\documentclass[aps,prb,eqsecnum]{revtex4}
\usepackage{amsmath,graphicx}
\usepackage{dcolumn}

\begin{document}

\title{The ontology of temperature in nonequilibrium systems}
\author{Alexander V. Popov}
\thanks{Permanent address: Technological Institute, Kemerovo 650056, Russia}
\author{Rigoberto Hernandez}
\thanks{Author to whom correspondence should be addressed}
\email{hernandez@chemistry.gatech.edu.}
\affiliation{Center for Computational and Molecular Science and Technology, \\
School of Chemistry and Biochemistry, \\
Georgia Institute of Technology, \\
Atlanta, GA  30332-0400}

\begin{abstract}
The laws of thermodynamics provide a clear concept of the temperature
for an equilibrium system in the continuum limit.
Meanwhile, the equipartition theorem allows one to make a connection between
the ensemble average of the kinetic energy and the uniform temperature.
When a system or its environment is far from equilibrium, however,
such an association does not necessarily apply.
In small systems, the regression hypothesis may not even apply.
Herein, we show that in small nonequilibrium systems, the regression
hypothesis still holds though with a generalized definition of
the temperature.
The latter must now be defined for each such manifestation.

\end{abstract}
    \maketitle

\section{Introduction}


A lay person defines temperature as a measure of the heat contained in a body.
This is not in conflict with a more rigorous thermodynamic definition
based on the equipartition theorem when the body
is in the continuum limit with respect to size and measurement time.
But what is the temperature when these assumptions no longer hold?
Is it useful to even try to describe small and/or nonequilibrium subsystems
using an instantaneous temperature?

Experimentally, at least, temperature can be defined even for
very fast and small systems using heat balances.
For example, the temperature of a medium
can be changed experimentally
using a system excitation induced by a laser pulse or
using a microwave field to heat the bath.
The temperature in the local environment can be raised quickly
by several to tens of degrees Centigrade
in a few microseconds or less.\cite{dyer00}
This elevated temperature
can be sustained for milliseconds and subsequently decays back to
the global environmental temperature within a time-scale
of tens of milliseconds.~\cite{kliger05}
Such $T$-jump experiments provide the time resolution needed for
studying the kinetics of many processes
---as in, {\it e.g.}, protein- and
peptide-folding,~\cite{snow04, kliger05, dyer00} and
helix-coil transitions in peptides.\cite{small00, eaton97, eaton97a, davis87}
However, $T$-jumps can significantly influence
the structural and dynamical properties of the systems under investigation.
For example, solutions of colloidal and microgel particles
change their size dramatically due to changes in temperature
---and also pH.\cite{lyon00,lyon01,lyon03c,nieves00,nieves01,nieves01a,fernan00,fernan03}
Indeed, such particles have been swelled by as much as an order of
magnitude in just a few milliseconds.\cite{pelton00,lyon01,fernan02}
Such environmental changes evidently affect the local and global structure
and may enable coupling to additional heat sinks.
At the very least, they require additional care in performing
the heat balance to determine the temperature during the $T$-jump.

In order to assign a rigorous definition of the instantaneous
temperature of a small (open or closed) system
under nonequilibrium conditions, it is useful to partition it
into a subsystem ---specifying the properties of the material---
and a bath ---consisting of everything else.
The latter may in turn also be open or closed depending on whether or
not it is effectively interacting with an additional environment.
When this bath is stationary,
such as can be found in simple liquids or colloidal suspensions at equilibrium,
the motion of the subsystem
---as represented by an $n$-dimensional position coordinate $q$---
can be accurately
described by the generalized Langevin equation (GLE),\cite{zwan01}
\begin{equation}
     \label{eq:GLE}
\ddot{q}(t)=-\frac{\partial V(q)}{\partial q}-\int_0^t{\gamma_{\rm
th}(t-t')\dot{q}(t')dt'}+\xi_{\rm th}(t)\;.
\end{equation}
Here $V(q)$ is the potential of mean force (PMF),
$\gamma_{\rm th}(t-t')$ is the friction kernel representing
the response of the solvent, $\xi_{\rm th}(t)$ is the
random force due to the medium, and they obey the
fluctuation-dissipation relation (FDR),
\begin{equation}
     \label{eq:FDR0}
\left<\xi_{\rm th}(t)\xi_{\rm th}(t')\right>
    =k_{\rm B}T\gamma_{\rm th}(t-t')\;.
\end{equation}
When the bath is nonstationary
by way of exhibiting temporal (and spatial) changes
in the ambient bath,
Eq.~(\ref{eq:GLE}) has been modified
to the so-called irreversible GLE (iGLE):\cite{hern99a,hern99e}
\begin{equation}
     \label{eq:iGLE}
\ddot{q}(t)=-\frac{\partial V(q)}{\partial q}
-\int_0^t{\gamma(t,t')\dot{q}(t')dt'}+\xi(t)\;,
\end{equation}
\begin{equation}
     \label{eq:KERNEL_iGLE}
\gamma(t,t')=g(t)\gamma_{\rm th}(t-t')g(t')\;.
\end{equation}
The function $g(t)$ modulating the friction kernel is completely
determined by an irreversible process due to processes not
otherwise included in the subsystem or bath.
The stochastic force $\xi(t)\equiv g(t)\xi_{\rm th}(t)$
in the iGLE is modulated by the same function $g(t)$.
The resulting nonstationary FDR is
\begin{equation}
     \label{eq:FDR_iGLE}
\left<\xi(t)\xi(t')\right>=k_{\rm B}T\gamma(t,t')\;.
\end{equation}


The system-bath dynamics characterized by the GLE and the iGLE
have often been generalized to
allow for the possibility of time-dependent or driven
variations in the system temperature.
In one set of approaches, the nonequilibrium behavior of the bath
is introduced by way of an external stochastic force.
This additional noise can drive the reaction coordinate $q$ directly\cite{ray00}
or by modulation of the local bath modes.\cite{sancho89,ray01,banik06}
Although such a modulation may be caused by many different
physically relevant mechanisms,
the recurring requirement is that the external noise has no connection to
the memory kernel $\gamma(t,t')$ in the FDR.
The latter can arise, for example, when the stochastic noise acting
on the reacting subsystem and the local bath is independent.
This requirement necessarily leads to a shift in the temperature.
Refs.~\onlinecite{sancho89, ray00, ray01, banik06} have
explored this issue within the Kramers theory, and
the steady state of the nonequilibrium open subsystem was observed
to lead to a new effective temperature.

However, if the system is truly in equilibrium with the overall bath,
then the heat transfer leading to this temperature differential is
presumably dissipated by bath modes not initially included in the
local bath.
A repartitioning of the environment into two sets
---a nonequilibrium open reservoir and a global thermally equilibrated bath---
have been suggested.\cite{MR95,ray98}
External perturbations in the initial state of the former
are relaxed to equilibrium through the coupling to the latter.
(The FDR between the random forces and the memory kernel appears
naturally in this case.)
This coupling is stipulated by a specific Hamiltonian model of the bath,
though primarily their work has specified the bath
using the usual choice of
a set of harmonic modes coupled by bilinear interactions.
Each such mode is then assumed to approach its equilibrium value
exponentially with a specified dissipation rate.
Alternatively, the heat from the nonequilibrium local bath may
be balanced directly in the iGLE through the
{\it ad hoc} introduction of an effective heat sink which has indeed
been seen to provide self-consistent constant-temperature
dynamics.\cite{hern02c, hern04c, hern05c}

Regardless, the requirement that the overall bath dissipate the
system so quickly that constant temperature is always maintained is
too severe.
For example, it does not
describe systems in which energy transport fluctuations
to nearby local bath modes may be important.
Nor does it account for
(driven) temperature-ramped chemical reactions
as has earlier been treated using a simple generalization of the
iGLE.\cite{hern99c}
That approach is restricted to the case of slow temperature
variations as will be recapitulated in Sec.~\ref{sec:discussion}.
Thus the question remains as to whether
a subsystem within a nonequilibrium environment can be characterized
using an observable temperature,
and if so, how to calculate or measure it.

\begin{figure}[ht]
\begin{center}
\includegraphics[width=8.0cm]{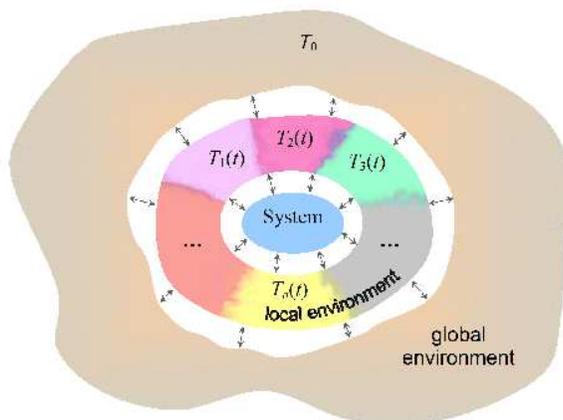}
\end{center}
\caption{A hierarchical representation of a (sub)system under
investigation which is immersed within a local nonequilibrium
environment that is, in turn,
interacting with a global equilibrium one.
(Color online.)} \label{fig:real}
\end{figure}

In the current work, we investigate a generalized chemical process
in which the nonequilibrium behavior of the chosen subsystem is
influenced by a change in temperature of the environment that is in
turn influenced by the response of the subsystem.
In such a case, the environment itself can also persist in a
nonequilibrium state.
The model, therefore, consists of a hierarchy of three domains:
a small-dimensional system,
a nonequilibrium open ``local'' bath (see Fig.~\ref{fig:real})
strongly coupled to the system,
and a ``global" bath that is coupled to the local bath and at most
weakly coupled to the system.
%
Without loss of generality, the 
linear response of the local baths can be
represented using an auxiliary model of 
harmonic oscillators coupled bilinearly to the system.
The interaction between the local and global environments induces
a time dependence in the parameters characterizing the local
bath---{\it viz.}, the effective masses of the oscillators,
their frequencies and the coupling between the local bath and the system.
Assuming that the effective temperature of each bath mode is observable,
we investigate the connections between these variables and the system.
The time dependence of the bath parameters can be specified according
to the behavior of any particular physical problem, and
so this theory can be applied quite generally as discussed above.


One important case for these problems (indeed it lies at the core of
the discussion that follows) is the possibility that the local bath
actually consists of several independent baths (or reservoirs).
This separation can be viewed as a mathematical trick that enables
one to demonstrate the possibility of invariances regardless of
the specific representation.
But it is also physically relevant.
For example, two- and three- temperature models have been used to
describe ultra-short laser pulse desorption
experiments.\cite{hohlfeld00,nest04}
Therein, the distinct reservoirs represent the adsorbate layer,
the substrate phonons and the substrate electrons, respectively,
and the times scales of each can be quite
different.\cite{saalfrank06}
However, these methods have typically been applied in an ad hoc
fashion, and the present work is an effort towards making this
theoretical structure clear in a general sense for cases such as this.

The GLE can be derived from a Zwanzig-type Hamiltonian as a
projection of the simplest many-dimensional mechanical system.\cite{zwan01}
Within this formalism, the bath is represented as a set of harmonic oscillators coupled
with a tagged particle by bilinear interactions.
In Ref.~\onlinecite{hern99e}, this formalism has been extended to
take into account nonstationary effects
by introducing
time-defendant coupling coefficients and a nonlocal
memory correction term (see Section~\ref{sec:DERIVATIONS}).
The projection of this system has been shown to be the
iGLE with the corresponding FDR.\cite{hern05b}
Although the form of the iGLE had been known earlier from the thermodynamic
considerations,\cite{hern99a}
the value of this approach is that it allows one to obtain
practical stochastic equations at a low cost.
For example, the local limit of the iGLE has been seen to surmise
the probe dynamics in
a nonstationary colloidal suspension.\cite{hern06e}

In Sec.~\ref{sec:DERIVATIONS}, we take advantage of the iGLE and its associated
Hamiltonian
to derive the stochastic equation
for a
tagged particle moving in a complex nonequilibrium environment consisting
of the reservoirs with changing temperatures.
We show that this approach can serve as a common basis to unify
various methods developed separately for specific purposes
in Sec.~\ref{sec:discussion}.
Moreover, a numerical gedenken experiment looking at the dynamics
of a particle coupled to three distinct baths (that are not mutually
coupled) at different temperatures illustrates the accuracy of the
nontrivial nonequilibrium effective temperature equation
derived in Sec.~\ref{sec:DERIVATIONS}.


\section{Langevin dynamics in a nonequilibrium environment}
                  \label{sec:DERIVATIONS}

\begin{figure}[ht]
\begin{center}
\includegraphics[width=8.0cm]{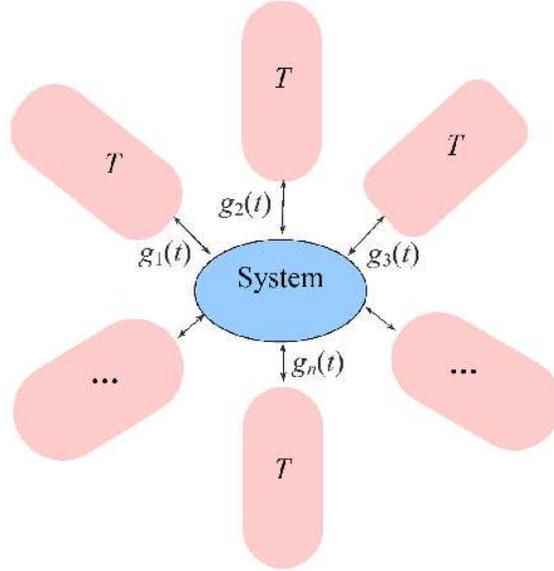}
\end{center}
\caption{A schematic representation of the generalized model
described in this work.
The subsystem interacts with multiple bath reservoirs by means of
time-dependent couplings while the temperature in each
bath is also subject to change.
(Color online.)}
\label{fig:model}
\end{figure}

The objective of this work is the determination of the
effective temperatures of the ``system'' shown
in Fig.~\ref{fig:model}.
The target system is directly coupled to a large number of
local reservoirs.
These reservoirs are, in turn, coupled to a larger scale environment
---to wit, the global environment---
that alters its properties and interaction with the system
over much longer time scales than the reservoir relaxation times.
In principle, a detailed Hamiltonian can be written that would include
the degrees of freedom for the system,
the local environments and the global environment illustrated
in Fig.~\ref{fig:real}.
Given such a specification, a projection of the Hamiltonian to the system
alone, and a study of its correlation functions would reveal its effective
temperature and related properties.
However, we found earlier that the projection onto a system coupled
to a single time-dependent local reservoir can be described by the
iGLE and an associated time-dependent Hamiltonian.\cite{hern99e,hern05b}
In what follows, we thus take a simpler approach to the solution of
our objective by generalizing the iGLE and its associated time-dependent
Hamiltonian for the case that the system can be coupled to
many distinct local reservoirs.
While much of the discussion takes advantage of the form of the
auxiliary Hamiltonian system to derive various correlation functions,
all of the latter involve only variables of the reduced dimensional system.
As such, these results are independent of the particular specification
of the auxiliary Hamiltonian.

\subsection{Multiple-reservoir model with fixed temperature}
                  \label{sec:FixedTMRmodel}

\begin{figure}[ht]
\begin{center}
\includegraphics[width=8.0cm]{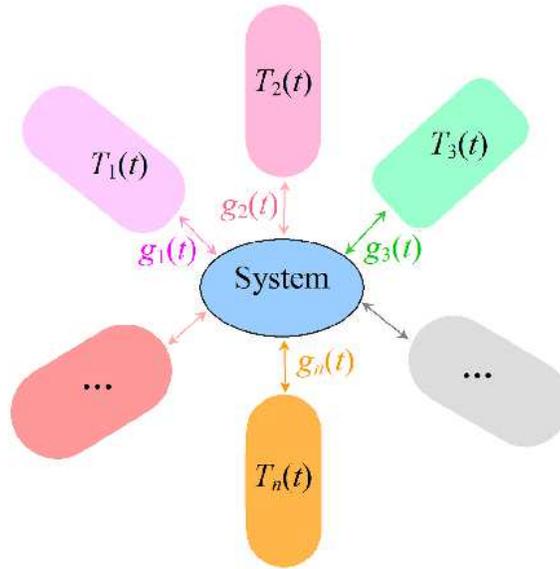}
\end{center}
\caption{A schematic representation of the
model investigated in
Refs.~\onlinecite{hern99a}, \onlinecite{hern99e}, and
\onlinecite{hern06e}.
The subsystem interacts with multiple bath reservoirs by means of
time-dependent couplings while the temperature remains constant.
(Color online.)}
\label{fig:Tconst}
\end{figure}

The Brownian motion of a tagged ---or chosen--- particle immersed in a
single equilibrium thermal bath is well understood.
More generally, however, the local environment of the particle
can be further divided into a set of
different reservoirs ---{\it viz.} distinct local baths---
with separately identifiable characteristic properties.
In this section,
all of these reservoirs are assumed to be at the same temperature
and therefore independently satisfy the same Boltzmann distribution
over their coordinates, ({\it c.f.}, Fig.~\ref{fig:Tconst}).
A more general case in which these reservoirs may
follow different temperatures
---and consequently, as a whole, the environment does not follow
a simple sing-temperature Boltzmann distribution---
is addressed in the next section.

In order to describe the dynamics of a particle solvated
by many distinct reservoirs,
one can generalize slightly the Zwanzig-type
Hamiltonian:\cite{hern99e,hern06e}
\begin{eqnarray}
                  \label{eq:HAM0}
H&=&\frac{p_q^2}{2} + V(q)
+ \delta V_1(q,t)
+ \delta V_2(q(\cdot),t)
\nonumber\\
&& -\sum_k g_k(t)\sum_ic^{(k)}_ix^{(k)}_iq
+\sum_k H^{(k)}_{\rm b}
\;,
\end{eqnarray}
where
each reservoir, $k$, is represented by a distinct set of harmonic modes.
The modes in each reservoir interact directly only with each other
and the tagged particle,
and indirectly to modes in other reservoirs  through
their mutual coupling to the tagged particle.
In the Hamiltonian of Eq.~(\ref{eq:HAM0}),
the sum of the first and second terms, $p_q^2/2+V(q)$,
is the bare energy of the tagged particle.
The last term is a sum of Hamiltonians,
\begin{equation}
                  \label{eq:HAM_BATH0}
H_{\rm b}^{(k)}=
 \frac{1}{2}\sum_i \left[\left(p^{(k)}_i\right)^2
 +\left(\omega_i^{(k)}x^{(k)}_i\right)^2\right] \;,
\end{equation}
each of which
represents the energy of the {\it k}-th bath reservoir with
coordinates $x_i^{(k)}$ and momenta $p_i^{(k)}$.
The fifth term provides
the bilinear coupling between the bath modes and the particle,
in which $\sum_k$ and $\sum_i$ denote the summation over the bath reservoirs
and the bath modes ({\it i.e.}, bath frequencies).
The third term,
\begin{equation}
                  \label{eq:V1}
\delta V_1(q,t)= \frac{1}{2}\sum_k\sum_i g_k^2(t)
\left(\frac{c^{(k)}_i}{\omega_i^{(k)}}\right)^2q^2
\;,
\end{equation}
is the renormalization of the potential which eliminates the
time-dependent spectral shift, and the fourth term,
\begin{equation}\label{eq:V2_0}
    \delta V_2(q(\cdot),t)=\frac{1}{2}\int_0^t dt' a(t,t')[q(t')-q(t)]^2 -
 \frac{1}{2}q(t)^2\int_0^t dt' a(t,t') \;,
\end{equation}
provides the time-dependent correction to the memory
effects.\cite{hern99e,hern06e}
The nonstationary memory correction in Eq.~(\ref{eq:V2_0}) was
defined earlier as
\begin{equation}
a(t,t')= \sum_k g_k(t)\dot g_k(t')\gamma_{{\rm (th)} k}(t-t')\;,
\end{equation}
where
\begin{equation}
     \label{eq:KERNEL_K}
\gamma_{{\rm (th)}k}(t-t')=\sum_i\left(\frac{c^{(k)}_i}{\omega_i^{(k)}}\right)^2
\cos\omega_i^{(k)}(t-t')
\end{equation}
is the friction kernel for the $k$-th reservoir at thermal equilibrium.

The nonlocality in the $\delta V_2(q(\cdot),t)$ term
has been addressed earlier\cite{hern99e,hern06e} and is
necessary because of the time dependence in the coupling terms.
(If said time dependence were zero then this term would also be zero.)
The former, in principle, also arises in
the case of space-dependent
friction\cite{lind81,carm82,carm83,lind83,lind84}
---{\it viz.} space-dependent coupling---
where the factors $g_k$ in the fifth
term of Eq.~(\ref{eq:HAM0}) depend only on the coordinate $q$.
Although the transient coupling terms
have recently been noted in said case,\cite{mahato95, mahato96}
they have been completely discarded in the corresponding
Hamiltonian equations of motion
by way of heuristic arguments concerned with the
possibility that they appear to violate causality.
In the present case, however,
where several coupling terms depend explicitly on time,
the nonstationary memory correction $\delta V_2(q(\cdot),t)$
---requiring the integration over the particle's path $q(\cdot)$
over all time--- cannot be discarded.
Indeed, this correction is necessary so as to
ensure that the projection onto the iGLE does not
itself contain additional nontrivial terms
as detailed in Appendix~\ref{app:B} that would violate causality.
As detailed earlier the component of the nonstationary memory
correction that violates causality does not present any problems because
it is smaller than other terms that have also been neglected
in this formalism.

In Ref.~\onlinecite{hern99e}, the iGLE
---Eq.~(\ref{eq:iGLE})---
(with a family of nonstationary friction kernels)
was shown to be  the projection of the equations of motion derived
from Hamiltonian~(\ref{eq:HAM0}) when there is only one reservoir.
In the current case of multiple reservoirs,
the friction kernel in the iGLE can be represented
as a sum of equilibrium friction kernels modulated by
distinct functions $g_k(t)$ defining the coupling strength
to the corresponding bath reservoirs,
\begin{equation}
     \label{eq:KERNEL_iGLE_MOD}
\gamma(t,t')=\sum_k g_k(t)g_k(t')\gamma_{{\rm (th)}k}(t-t')\;.
\end{equation}
The functions $g_k(t)$ are specified by the time-dependence of the coupling
up to a trivial arbitrariness in their initial values that is renormalizable.
In cases when the reservoir is initially uncoupled to the particle,
then said values are necessarily specified and equal to zero, however.

The FDR~(\ref{eq:FDR_iGLE}) for the projection of the iGLE,
Eq.~(\ref{eq:iGLE}), with the multi-reservoir friction,
Eq.~(\ref{eq:KERNEL_iGLE_MOD}),
results from the Hamiltonian of Eq.~(\ref{eq:HAM0}),
if the initial distribution is of the form,
\begin{eqnarray} \label{eq:P}
P&\propto& \exp
\left[-\frac{1}{2k_{\rm B}T}\sum_{k,i}
  \left(
     \left[ p^{(k)}_i(0)\right]^2
\right.\right.\nonumber\\&&\left.\left.
+\left[ \omega_i^{(k)}x^{(k)}_i(0)-\frac{c^{(k)}_ig_k(0)}{\omega_i^{(k)}}q(0)
        \right]^2
  \right)
\right]\;,
\end{eqnarray}
and the conditions
\begin{subequations}\begin{eqnarray}
                  \label{eq:AVER}
&&\left<p^{(k)}_i(0)p^{(l)}_j(0)\right>=k_{\rm B}T\delta_{ij}\delta_{kl}\;, \\
                  \label{eq:AVER1}
&&\left<\left(\omega_i^{(k)}x^{(k)}_i(0)-\frac{c^{(k)}_ig_k(0)}{\omega_i^{(k)}}q(0)\right)
\right.\nonumber\\
 && \left.\times
\left(\omega_j^{(l)}x^{(l)}_j(0)-\frac{c^{(l)}_jg_l(0)}{\omega_j^{(l)}}q(0)\right)\right>
\nonumber\\ && =
k_{\rm B}T\delta_{ij}\delta_{kl}\;,\\
                  \label{eq:AVER2}
&&\left<p^{(k)}_i(0)\left(\omega_j^{(l)}x^{(l)}_j(0)
-\frac{c^{(l)}_jg_l(0)}{\omega_j^{(l)}}q(0)\right)\right>
\nonumber\\ && = 0
\end{eqnarray}\end{subequations}
are satisfied.

The iGLE with multiple and distinct reservoirs has thus been specified,
and allows for temporal changes in the response of the environment albeit
while all such reservoirs maintain the same constant temperature.


\subsection{Multiple-reservoir model with time-dependent response}

\subsubsection{Squeezing bath modes}

In principle, the effective temperature of each of the reservoirs
(solvating the tagged particle)  need not be the same
as  they are not coupled to each other directly.
This situation is illustrated in
Fig.~\ref{fig:model} in which the temperature of each reservoir
is different, perhaps as a consequence of some unspecified
external or internal processes.
It can be implemented within the Hamiltonian representation of
a particle interacting with multiple reservoirs
by supposing that the bath modes are not
static---that is, their effective masses and coordinate
scales are allowed to vary in time.

The time-dependent functions,
$h_k(t)$ and $\mu_k(t)$,
are now introduced so as to represent the time-dependent
scaling in the force constant and mass of the $k^{\rm th}$
bath mode, respectively.
The Hamiltonian of the $i$-th oscillatory bath mode belonging to
the $k$-th reservoir can thereby be written as
\begin{equation}
                  \label{eq:HAM_BATH}
H^{(k)}_{{\rm b},i}(x^{(k)}_i,p^{(k)}_i,t)= \frac{\left[p^{(k)}_i\right]^2}{2\mu _k^2(t)}
+\frac{1}{2}\left[\omega_i^{(k)}h_k(t)x^{(k)}_i\right]^2\;.
\end{equation}
The terms, $h_k(t)$ and $\mu_k(t)$, combine to
determine the time dependence of the mode
frequencies in the {\it k}-th reservoir.
Their initial values can be chosen as
$h_k(0)=\mu_k(0)=1$ without loss of generality.
Meanwhile, the scaling of the force constant effectively narrows
(or widens) the configuration space available to the oscillator
at a given temperature.
Such a trick of ``squeezing" bath oscillators
leads to changes in the total energy of the reservoir,
and thereby shifts the reservoir temperature.
The existence of such temperatures relies on the fact that
the relaxational redistribution of the collective energy
within the modes is sufficiently fast that bath reequilibrates
on the time scales of the particle motion.
In other words, the process of bath mode squeezing is assumed to be
adiabatic and to vary on time scales longer than the
periods of oscillation in the nontrivial modes of a given reservoir.

The time-dependent Hamiltonian~(\ref{eq:HAM0}) can now be
modified by introducing the scaling
functions $h_k(t)$ and $\mu_k(t)$:
\begin{eqnarray}
                  \label{eq:HAM}
H&=&\frac{p_q^2}{2} + V(q)
 + \delta V_1(q,t)
 + \delta V_2(q(\cdot),t)
\nonumber\\
 &-& \sum_{k,i} c^{(k)}_ig'_k(t)h_k(t)x^{(k)}_iq
 + \sum_{k,i} H^{(k)}_{{\rm b},i}(x^{(k)}_i,p^{(k)}_i,t)
\;.\nonumber\\
\end{eqnarray}
Here the coupling coefficients $g'_k(t)$ are introduced so as to
simplify the structure of the Hamiltonian.
These coefficients are unequal to the earlier coefficients $g_k(t)$.
Nevertheless they can be connected through the expression,
\begin{equation}
                  \label{eq:new_g}
g'_k(t)=g_k(t)\sqrt{\frac{h_k(t)}{\mu_k(t)}} \;,
\end{equation}
allowing comparison to the earlier results obtained using
the original iGLE.\cite{hern99a,hern99c,hern99e,hern06e}
In Eq.~(\ref{eq:HAM}) the renormalization potential, $\delta V_1$, is given
by Eq.~(\ref{eq:V1}) with
$g_k(t)$ replaced by $g'_k(t)$.
The nonstationary memory correction, $\delta V_2$,
is defined by Eq.~(\ref{eq:V2_0}) with
\begin{subequations}\label{eq:A_NONLOC}
\begin{equation}
a(t,t')= \sum_k \frac{h_k(t)}{\mu_k(t)} g_k(t)\dot g_k(t')
\gamma_{{\rm(th)}k}\left[\tau_k(t)-\tau_k(t')\right]\;,
\end{equation}
where $g_k(t)$ has not been replaced by $g'_k(t)$,
\begin{equation}
\tau_k(t)\equiv \int_0^t h_k(s)/\mu_k(s)ds
\end{equation}
and $\gamma_{{\rm(th)}k}(\cdot)$ is defined in Eq.~(\ref{eq:KERNEL_K}).

\end{subequations}
Note that the coupling coefficients $g_k(t)$ can also depend on the temperature.
Although in the framework of this approach,
there is no recipe for connecting the
temperature-driving functions $\mu_k(t)$ and $h_k(t)$ to these coefficients,
such a connection can in principle be obtained from a deeper analysis
based on mode-coupling theory, direct extraction of these functions
from simulations (Ref.~\onlinecite{hern06e}; see also Sec.~\ref{sec:stochastic}
below), or experiments.


As was done in the earlier section with respect to the behavior of $g_k(t)$,
the functions $\mu_k(t)$ and $h_k(t)$ must be required to be
effectively stationary on the correlation time scales of the
fastest bath
modes.\cite{MR95,ray98}
This so-called slow-bath-mode ``squeezing" assumption
is necessary because otherwise there would always exist a bath mode
whose response to fluctuations would be so chaotic that it wouldn't
ever relax to quasi-equilibrium limits.
Evidently, for low spectral frequencies, this imposes some restrictions
on the rates of change of $\mu_k(t)$ and $h_k(t)$
and, thus, implies a related assumption
that low frequency bath modes contribute negligibly
to the particle's motion.
These assumptions are valid, for example,
in the over-damped case when the spectral density, $J(\omega)$,
of the bath Hamiltonian is proportional to $\omega^s$ with $s > 1$
at $\omega \rightarrow 0$, and hence the
small bath frequencies are not coupled to the system.

%

\subsubsection{The effective temperature of each reservoir}

The strategy for solving the dynamics of Eq.~(\ref{eq:HAM_BATH}),
taking into account its hierarchical structure, focuses first on the
solution of the local dynamics of the bath modes within a given
$k^{\rm th}$ reservoir as if it does not interact with the rest of the modes.
In the next section, we will use these states as a reference for
the approximate solution of the global dynamics.
The accuracy of this approach relies on the assumption that the
coupling between the reservoirs and the tagged particle is weak,
as has indeed been assumed from the outset.
For simplicity, in this subsection the upper indices $(k)$ will often
be omitted as all of the formulas herein are necessarily being solved within
a given $k^{\rm th}$ reservoir.

At any given time, the energy manifold of the $i^{\rm th}$ bath
mode in the $k^{\rm th}$ reservoir is simply a function of the
phase space variables, $(x_i,p_i)$, determined by the instantaneous
parameters, {\it i.e.},
\begin{equation}\label{eq:ikHam}
E_i(t)=\frac{p_i^2}{2\mu_k^2(t)}
      +\frac{\omega_i^2}{2}h_k(t)^2x_i^2
\;.
\end{equation}
If $\mu_k(t)$ and $h_k(t)$ change slowly relative to the oscillator frequency,
then the symplectic area of this manifold ---{\it viz.}, the action $I$---
will be an adiabatic invariant.
On the elliptical energy manifold defined by Eq.~(\ref{eq:ikHam}), this
invariant is given by
\begin{equation}
I =\frac{\mu_k(t)E_i(t)}{h_k(t)\omega_i}
  =\frac{\mu_k(0)E_i(0)}{h_k(0)\omega_i}
  =\frac{E_i(0)}{\omega_i}
\;,
\end{equation}
where the last equality follows from the initial choice,
$h_k(0)=\mu_k(0)=1$.
The time-invariance of $I$ thus provides a useful connection,
\begin{equation}\label{eq:Eit-Ei0}
E_i(t)=E_i(0)\frac{h_k(t)}{\mu_k(t)}\;,
\end{equation}
between the energies of the $i^{\rm th}$ bath mode at various times.

Given that each bath mode is initially equilibrated at its
corresponding reservoir temperature $T_k(0)$
---{\it i.e.}, the initial energies $E_i(0)$ of the reservoir
are Boltzmann distributed,---
the connection in Eq.~(\ref{eq:Eit-Ei0}) readily implies that
the energies $E_i(t)$ at any given time are also Boltzmann
distributed at the temperature,
\begin{equation}
                  \label{eq:T_OF_t}
T_k(t)=T_k(0)\frac{h_k(t)}{\mu_k(t)}\;.
\end{equation}
Thus, the ratio $h_k(t)/\mu_k(t)$ establishes
the time dependence in the temperature
of the $k^{\rm th}$ bath reservoir.
Although the interaction between the reservoir and the tagged particle
could, in principle, alter this structure by way of heat transfers,
in practice it does not do so.
The bath reservoir is assumed to be
sufficiently large that such transfers are small in comparison with the
total energy of the reservoir.
(This assumption is not very severe because it must be satisfied in
order for the bath to have a well-defined quasi-equilibrium temperature!)
The functions $h_k(t)$ and $\mu_k(t)$ are auxiliary and
are not known {\it a priori},
but the time-dependent temperature entering in
Eq.~(\ref{eq:T_OF_t}),
together with the coupling strength coefficients $g_k(t)$,
can either be found from simulations or estimated from experimental data.
For example, they have been obtained in the case of a diffusing particle
in a swelling colloidal suspension in Ref.~\onlinecite{hern06e}.


The initial temperatures of the various bath reservoirs need not be
the same because some modes may be connected to distinct heat sources
and sinks leading to steady state energy transport
through the system between the reservoirs.
Indeed, the nonequilibrium dynamics of a particle connected
to reservoirs with different time-independent temperatures
has been explored by Kurchan and coworkers.\cite{kurchan00,kurchan05}
But if the system is initially in equilibrium, then
in the absence of such energy flows, the initial temperatures of all the
reservoirs will be the same.
More generally, however, the initial probability distributions
---cf.~Eq.~(\ref{eq:P})---
for a given reservoir can be rewritten as,
\begin{eqnarray}
                  \label{eq:P_new}
P_k&\propto& \exp
  \left\{ -\sum_{i}\frac{1}{2k_{\rm B}T_k(0)}
    \left(
       \left[p^{(k)}_i(0)
       \right]^2
\right.\right.\nonumber\\ && \left.\left.
      +\left[ \omega_i^{(k)}x^{(k)}_i(0)-\frac{c^{(k)}_ig_k(0)}{\omega_i^{(k)}}q(0)
       \right]^2
     \right)
  \right\}\;,
\end{eqnarray}
for the $k^{\rm th}$ reservoir temperature, $T_k(0)$.
The correlations in Eqs.~(\ref{eq:AVER})-(\ref{eq:AVER2})
are still satisfied if one replaces the temperature, $T$
with that of the given reservoir, $T_k(0)$.
Such an initial equilibration seems to be artificial when the
temperatures $T_k$ change with time due to the interactions of reservoirs
with the tagged subsystem and the global bath.
Note, however, that the initial time, $t=0$, can be formally shifted
to the far past.
The system then has no memory of the initial conditions and, thus,
the resulting iGLE remains unaffected.

\subsubsection{Equations of motion}

The equations of motion for the many-reservoir case follows directly
from the Hamiltonian in Eq.~(\ref{eq:HAM}):
\begin{subequations} \label{eq:MOTION}
\begin{eqnarray}\label{eq:MOTION1}
\dot x_i^{(k)} &=& \frac{p_i^{(k)}}{\mu_k^2(t)}\;, \\
                  \label{eq:MOTION2}
\dot p_i^{(k)}
       &=& -\left[
               h_k(t)\omega_i^{(k)}
            \right]^2 x_i^{(k)}
          + c_i g'_k(t)h_k(t)q\;, \\
                  \label{eq:MOTION3}
\dot q   &=& p_q\;, \\
                  \label{eq:MOTION4}
\dot{p}_q &=& -\frac{\partial V(q)}{\partial q}
              +\sum_{k,i} c_i^{(k)} g'_k(t)h_k(t)x_i^{(k)} \\
           && - \sum_{k,i} \left(
\frac{c_i^{(k)}g'_k(t)}{\omega_i^{(k)}} \right)^2 q -
 \frac{\partial}{\partial q(t)}\delta V_2(q(\cdot),t)
\;, \nonumber
\end{eqnarray}
\end{subequations}
where the indices, $k$, referring to a given reservoir have been explicitly
retained to help the reader.
However, they are somewhat cumbersome,
and so in this section it is convenient to replace
explicit reference to the superscripted reservoir index.
Instead the collective index ${\bf i} (\equiv \{i,k\})$ will be used for
this reference, as in {\it e.g.} $\omega_{\bf i} \equiv \omega_i^{(k)}$.

As discussed in Sec.~\ref{sec:FixedTMRmodel},
the presence of the nonstationary memory correction
$\partial(\delta V_2)/ \partial q(t)$
in Eq.~(\ref{eq:MOTION})
introduces the possibility of violations of the
causality principal.
Indeed, the variational principle applied to the calculation of this term
gives an apparent contribution to the force from future trajectories.
However the error in this contribution is on the order of the
neglected terms in the perturbative treatment in the projection,
and it was seen earlier \cite{hern06e}
that all of these small errors can be safely neglected.
But the nonstationary memory correction
$\partial(\delta V_2)/ \partial q(t)$ can not be entirely neglected
because it contains a nontrivial correction
which ensures that the projection to the
iGLE does not violate causality up to the second order
of the perturbation expansion (c.f., Appendix~\ref{app:B}).

From Eqs.(\ref{eq:MOTION1}) and (\ref{eq:MOTION2}) one can derive
\begin{equation} \label{eq:MODE}
\ddot x_{\bf i}
   + \nu_k(t) \dot x_{\bf i}
   + \Omega^2_{\bf i}(t) x_{\bf i}
   = c_{\bf i} \frac{g'_k(t)h_k(t)}{\mu_k^2(t)}q \;,
\end{equation}
where
\begin{subequations} \label{eq:MODE_COEFF}
\begin{eqnarray}
\nu_k(t)
        &=& \frac{2\dot{\mu}_k(t)}{\mu_k(t)} \;,\\
\Omega_{\bf i}(t)
        &=& \frac{h_k(t)}{\mu_k(t)}\omega_{\bf i} \;.
\end{eqnarray}
\end{subequations}
The assumption of a slow``squeezing" bath modes, discussed before,
must now be restated in terms of both time dependent parameters,
$h_k(t)$ and $\mu_k(t)$.
After a little algebra, it can be shown to reduce to the requirement
\begin{equation}
\left| \frac{\dot{\mu}_k}{\mu_k} +
\frac{\dot{h}_k}{h_k} \right| \ll \Omega_{\bf i} \;,
\end{equation}
which must be satisfied for all modes $i$ in a $k^{\rm th}$ reservoir.
Under this assumption,
one obtains the following result
[as is shown in Appendix~\ref{app:A})]:
\begin{eqnarray}
                  \label{eq:Xi}
x_{\bf i} &=& u_{\bf i}(t)x_{\bf i}(0)
      + u_{\bf i}(t)\dot x_{\bf i}(0)
            \int_0^t \frac{dt'}{(u_{\bf i}(t')\mu_k(t'))^2} 
      \nonumber\\
      &+& c_{\bf i}u_{\bf i}(t)
          \int_0^t\frac{dt'}{(u_{\bf i}(t')\mu_k(t'))^2}
          \int_0^{t'}{g'_k(s)h_k(s)u_{\bf i}(s)q(s)ds}
\;, \nonumber\\
\end{eqnarray}
where the bath mode oscillations are described by
\begin{equation}
                  \label{eq:Ui}
u_{\bf i}(t)=\frac{\cos \omega_{\bf i}\tau_k(t)}{\sqrt{\mu_k(t)h_k(t)}} \;,
\end{equation}
and the effective time $\tau_k$ can be written as
\begin{equation}
                  \label{eq:TAU}
\tau_k(t) \equiv \frac{1}{\omega_{\bf i}}\int_0^t{\Omega_{\bf i}(t')dt'}=
\int_0^t{\frac{h_k(t')}{\mu_k(t')}dt'}=\int_0^t{\frac{T_k(t')}{T_k(0)}dt'} \;.
\end{equation}
Perhaps not surprisingly,
the effective time changes ``faster" when the temperature increases.
A similar transformation of time has been found earlier in
simplifying overdamped Langevin equations driven by
time-dependent temperature baths.~\cite{Reim02, hanggi81, reim96}

The projected equation of motion for the tagged particle
reduces to the iGLE form after the appropriate averaging over the
distributions derived in the previous subsection.
(The details of this derivation can be found in Appendix~\ref{app:B}.)
The nonstationary friction has a form similar to that
of Eq.~(\ref{eq:iGLE}) with
\begin{subequations}\label{eq:NSfriction}\begin{eqnarray}\label{eq:FRICTION}
\gamma(t,t')&=&
\sum_k \frac{T_k(t)}{T_k(0)}
g_k(t)g_k(t')\gamma_{{\rm(th)}k}(\tau_k(t)-\tau_k(t'))  \\
&=& \sum_k \sqrt{\frac{T_k(t)}{T_k(t')}} \gamma_k(t,t')
\end{eqnarray}
for $t>t'$,
where the weighted friction of the $k^{\rm th}$ mode is defined as
\begin{equation}
\gamma_k(t,t') \equiv
   \frac{\sqrt{T_k(t)T_k(t')}}{T_k(0)}
g_k(t)g_k(t')\gamma_{{\rm(th)}k}(\tau_k(t)-\tau_k(t'))
\;,
\end{equation}\end{subequations}
and is symmetric in the two times, $t$ and $t'$.
(Note that $\gamma_{{\rm(th)}k}(\cdot)$ is an even function
of its argument as can be seen in Eq.~(\ref{eq:KERNEL_K}).)
Recalling the definition for the bath reservoir temperature and friction kernel
in Eqs.~(\ref{eq:T_OF_t}), and Eq.~(\ref{eq:KERNEL_K}), respectively,
the nonstationary stochastic force is found to be
\begin{eqnarray}
\xi(t)&=&\sum_k\frac{T_k(t)}{T_k(0)}g_k(t)
\sum_i \frac{c_{\bf i}}{\omega_{\bf i}}
\nonumber\\
&& \times\left[
    p_{\bf i}(0)\sin\omega_{\bf i}\tau_k(t)
   + \left( \omega_{\bf i}x_{\bf i}(0)-\frac{c_{\bf i}g_k(0)}
            {\omega_{\bf i}}q(0)
     \right)
     \cos\omega_{\bf i}\tau_k(t)
   \right]
\;. \label{eq:FORCE}
\end{eqnarray}
Imposing the initial condition in Eq.~(\ref{eq:P_new}), the random force
correlation function becomes
\begin{subequations}\label{eq:XI_AVER}\begin{eqnarray}
\left<\xi(t)\xi(t')\right> &=& \sum_k
k_{\rm B}\frac{T_k(t)T_k(t')}{T_k(0)}
g_k(t)g_k(t')\gamma_{{\rm(th)}k}(\tau_k(t)-\tau_k(t')) \\
&=& k_{\rm B}\sum_k \sqrt{T_k(t)T_k(t')} \gamma_k(t,t')
\;.
\end{eqnarray}\end{subequations}

Equations (\ref{eq:NSfriction}) and (\ref{eq:XI_AVER})
can be used to recast the extended form of the FDR,
\begin{subequations}\label{eq:central}
\begin{equation}
                  \label{eq:XI_AVER_Teff}
\left<\xi(t)\xi(t')\right>= k_{\rm B}T_{\rm eff}(t,t')
\gamma(t,t')\;,
\end{equation}
where the effective temperature is
\begin{equation}
T_{\rm eff}(t,t') =
  \frac{\sum_k \sqrt{T_k(t)T_k(t')} \gamma_k(t,t')}%
       {\sum_k \sqrt{\frac{T_k(t)}{T_k(t')}} \gamma_k(t,t') }\;,
\label{eq:Teff}
\end{equation}
\end{subequations}
for $t>t'$.
Eqs.~(\ref{eq:central}) comprise the central result of this work
and generalize the FDR by including the simultaneous
interactions of the tagged particle
with different nonstationary thermal reservoirs.
In particular Eq.~(\ref{eq:Teff}) provides a
clear prescription of the effective temperature of the chosen particle
that emerges as a consequence of averaging the microscopic dynamics
of many reservoirs.
This is a nontrivial result because it dances between microscopic and
macroscopic quantities.
Moreover, as written, the effective temperature appears to be asymmetric
between the two times, $t$ and $t'$, and hence may cause alarm.
However, only the product $T_{\rm eff}$ and $\gamma$
in Eq.~(\ref{eq:XI_AVER_Teff})
must remain symmetric.
As shown in the discussion that follows, this product
is indeed symmetric with respect to
the two times in its argument,
and $T_{\rm eff}$ reduces to forms that have earlier been derived
or assumed for a number of limiting cases.

One possible concern with the underlying equations that have
led to the Eq.~(\ref{eq:central}) lies in the specification of the
initial condition for the system and its environment, given that there
must evidently be some environmental memory of the previous motion by way
of the values of $q(t)$ at earlier times.
In past and present work, we have
typically employed one of three boundary conditions:
{\it (i)} The system is at equilibrium at the initial time, 0,
just before the nonequilibrium disturbance is effected.
This requires us to run dynamics
starting at a time well prior to $t=0$ allowing the system to equilibrate
and fully prepare its environmental memory.  (This evidently requires
sampling over many trajectories.)
{\it (ii)} Different reservoirs at $t=0$ are at distinct equilibrium within
themselves, but are not at equilibrium with each other.  In this case, we
prepare each reservoir separately using the procedure described for
case {\it i}.
{\it (iii)} The particle is suddenly placed inside the reservoirs,
and hence there is
no prior memory due to motion before $t=0$.
In this section, we have applied the boundary conditions from
case {\it ii}, in
order to simplify the expressions from correction terms that would have
been trivial anyway.
Indeed, the use of case {\it iii} for the boundary conditions in
the numerical simulations shown below did not affect the results.


%
%
%


It may appear that the terms in the central result of Eq.~(\ref{eq:central})
are contingent on the creation of an associated Hamiltonian of the form 
of Eq.~(\ref{eq:HAM_BATH}) used in its analytical derivation.
However, as in other such reconstructions using the 
latter auxiliary infinite-dimensional representation, the projected form
of the central result contains no object with explicit reference to any
unprojected dynamics.
As such, it should be capable of describing any dynamics which is captured
by the continuum representation of the iGLE.
More specifically, the temperatures $T_k$ can be calculated or measured
directly from each of the independent baths.
The coupling terms $g_k$  and memory times $\tau_k$
can be calculated or measured directly by
obtaining the correlation functions of the forces exerted on the tagged
particle by each bath, respectively, upon removal of the thermal component.
Such a procedure has been used in Ref.~\onlinecite{hern06e} in the 
limit in which the chosen particle is coupled to only one bath.
Several additional examples are provided below
in which the particle is coupled to multiple baths.
In each of these cases, the associated functions, 
$g_k(t)$, $\tau_k(t)$, and $T_k(t)$, necessary for the application
of Eq.~(\ref{eq:central}) are constructed.
It is notable, that these functions can be obtained without recourse
to the auxiliary infinite-dimensional form of Eq.~(\ref{eq:HAM_BATH}).

\section{Results and Discussion of Limiting Cases} \label{sec:discussion}

In order to better understand the effective temperature expression
derived above, it is helpful to look at a number of limiting cases.
In Sec.~\ref{subsec:tdinhomoT}, the structure of the correlation
functions when
all the reservoirs are held at constant but distinct temperatures
is shown to agree with the earlier perspective of
Kurchan and coworkers.\cite{kurchan05}
In Sec.~\ref{subsec:tdhomoT}, the generalized Langevin equations used
earlier to describe aging systems are recovered for when the
reservoirs are all held at a homogeneous, but time-dependent, temperature.
This latter limit is also equivalent to that suggested earlier in
Ref.~\onlinecite{hern06e} in a case studying driven nonequilibrium colloidal
suspensions.
In Sec.~\ref{subsec:TwoTRes}, the two-temperature model of
Refs.~\onlinecite{MR95} and \onlinecite{ray98} is also recovered
in the limit that the system is coupled to two local reservoirs.
A set of gedenken experiments wherein the bath is coupled to
three local reservoirs is shown in Sec.~\ref{subsec:ThreeTRes}
to lead to a nontrivial effective temperature in agreement with the
central results of this work.
In Sec.~\ref{subsec:LocalQuasiEq}, the local quasi equilibrium limit
of a non-Markovian Fokker-Planck equation
explored earlier by Rubi and coworkers\cite{rubi04} is shown to
lead to correlation functions equivalent to those in the
corresponding limit of this work.
Thus the central result for the effective temperature of the system
driven by nonequilibrium local environments is shown to capture all
the previous limits as well as extend the formalism to heretofore
unknown cases.

Before proceeding, it is perhaps also useful to recapitulate the
separation of time scales in the hierarchy illustrated by Fig.~\ref{fig:real}
that has been assumed in constructing the central result in
Eq.~(\ref{eq:central}).
The relaxation times for responses to large perturbations
of the system $\tau_{\rm s}$, the local reservoirs
$\tau_k$, and the global environment $\tau_{\rm g}$ must satisfy
the simple ordering, $\tau_{\rm s} \le \tau_k \le \tau_{\rm g}$.
Meanwhile, the system is so small compared to the local reservoir
that any perturbation at times on the order of $\tau_{\rm s}$ 
leads to no effective change of the local reservoir behavior.
Any sustained change in the system behavior over times longer than
$\tau_{k^{'}}$ will affect and change the behavior of the 
${k^{'}}^{\rm th}$ reservoir, but it will simply lead to a change of the
quasi-static response of said reservoir to the system at the
short times scales near $\tau_{\rm s}$.
An analogous description is true for the time scales between the
local reservoirs and the global bath.
Because of these inequalities, any time-dependence in $T_k$ and
$g_k$ must be slow in comparison with the motion of the system.
However, as we have seen in earlier work, and shown here in the numerical
work, the time scales do not have to be nearly so disparate as the corrections
are small.

%
%

\subsection{Environments with Constant Temperature Inhomogeneities}
\label{subsec:tdinhomoT}

The structure of Eq.~(\ref{eq:XI_AVER}) and the associated stochastic
equations derived from them allows one to describe the dynamics
of a chosen particle connected
to many reservoirs where each can be at a different temperature.
If these temperatures are constant in time, {\it i.e.},
$T_k(t)=T_k(t')=T_k$, then from
Eqs.~(\ref{eq:TAU}), (\ref{eq:FRICTION}) and (\ref{eq:XI_AVER})
one obtains
\begin{eqnarray}
                  \label{eq:FRICTION_SWITCH}
\gamma(t,t')&=&
\sum_k
g_k(t)g_k(t')\gamma_{{\rm(th)}k}(t-t') \;, \\
                  \label{eq:XI_AVER_SWITCH}
\left<\xi(t)\xi(t')\right>&=& \sum_k
k_{\rm B}T_k
g_k(t)g_k(t')\gamma_{{\rm(th)}k}(t-t') \;.
\end{eqnarray}
An analogous case has been discussed in Ref.~\onlinecite{kurchan05} in the
stationary limit when $g_k(t)\equiv 1$ (stable non-uniform environment).
Thus, Eqs.~(\ref{eq:FRICTION_SWITCH}) and (\ref{eq:XI_AVER_SWITCH})
extend the previous treatment so as to include nonequilibrium baths.

The implementation of the iGLE theoretical framework discussed
above follows readily for any given nonequilibrium system interacting
with many nonequilibrium baths.
The key step is the identification of which baths
---{\it viz.}, temperature heat sinks---
are in contact with the chosen particle as a function of time.
Nonzero and zero values of $g_k(t)$ at any given time correspond
to whether said reservoir is connected and disconnected, respectively,
from the particle at the given time, $t$.
(An analogous procedure was used in Ref.~\onlinecite{jarz00}
for deriving a variant of the fluctuation theorem.)
This will help to switch the chosen subsystem between different
heat reservoirs with different temperatures.
If the reservoir temperature does not change
significantly during a period when an appropriate coupling term differs
from zero, one obtains Eqs.~(\ref{eq:FRICTION_SWITCH}) and
(\ref{eq:XI_AVER_SWITCH}).

\subsection{Environments with Time-Dependent Homogeneous Temperature}
\label{subsec:tdhomoT}

If the temperature changes uniformly for all the bath modes,
$T_k(t)=T(t)$, then Eq.~(\ref{eq:XI_AVER}) simplifies to
\begin{equation}
                  \label{eq:GFDR1}
\left<\xi(t)\xi(t')\right>= k_{\rm B}T(t')\gamma(t,t') \;.
\end{equation}
%
Such generalized FDRs (GFDRs) have been used
in the description of the nonequilibrium steady-state dynamics
of glassy systems, for example.\cite{ritort03, frey05c}
The relations between autocorrelation and response functions,
like that of Eq.~(\ref{eq:GFDR1}),
are frequently considered in the literature as a signature of
FDR violations (quasi-FDR) arising from partial equilibration among a
subset of degrees of freedom, and the quantity $T(t')$ is treated as an
effective temperature.
In the specific case of aging systems,\cite{ritort03, sollich02a, sollich02b}
the effective temperature depends on both $t$ and $t'$.
The name ``quasi-FDR" and the use of $T_{\rm eff}$ as an auxiliary parameter
are reasonable due to the fact that the actual thermodynamic temperature
is constant in these studies.

The results of the previous section, however, allow for the recognition that
$T(t')$ is the true ambient temperature at the time $t'$,
and the interpretation that GFDR in Eq.~(\ref{eq:GFDR1})
expresses the transient behavior of the system.
Note, that in the iGLE model, the partial relaxation of the baths
corresponds to
different dependencies $T_k(t)$
[included into the GFDR, Eq.~(\ref{eq:XI_AVER})],
and leads to the two-time dependence of $T_{\rm eff}$ in
Eq.~(\ref{eq:XI_AVER_Teff}).

In the specific case when all the coupling coefficients are also
equal, $g_k(t)=g(t)$, we obtain
\begin{equation}
                  \label{eq:FRICTION2}
\gamma(t,t')=\frac{T(t)}{T(0)}g(t)g(t')\gamma_{\rm th}(\tau(t)-\tau(t')) \;,
\end{equation}
for $t>t'$, where
$ \gamma_{\rm th}(\tau-\tau') \equiv
  \sum_k \gamma_{{\rm(th)}k}(\tau-\tau')$
is the friction kernel of the whole system at equilibrium, and
\begin{subequations}\label{eq:FORCE1}
\begin{equation}
\xi(t)= \frac{T(t)}{T(0)}g(t)\xi_{\rm th}(\tau(t))\;,
\end{equation}
with
\begin{equation}
\left<\xi_{\rm th}(\tau)\xi_{\rm th}(\tau') \right>=k_{\rm B}T(0)
\gamma_{\rm th}(\tau-\tau') \;.
\end{equation}
\end{subequations}

It is instructive to compare these results with those obtained in
Ref.~\onlinecite{hern99c} from phenomenological considerations.
Namely, the friction kernel used in that work,
\begin{equation}
                  \label{eq:FRICTION_RH}
\gamma(t,t')=g(t)g(t')\gamma_{\rm th}(t-t')\;,
\end{equation}
coincides with Eq.~(\ref{eq:KERNEL_iGLE}) and may include
effects due to the temperature change implicitly through the
coupling coefficient $g(t)$ only.

This restriction on the memory kernel also induces the main
difference between the two approaches: the use of the real time $t$
{\it vs.} the effective one, $\tau(t)$.
Hence, only the amplitude of the stochastic force changes along with the
temperature alteration, and gives
\begin{equation}
                  \label{eq:FORCE_RH}
\xi(t)= \sqrt{\frac{T(t)}{T(0)}}g(t)\xi_{\rm th}(t)\;.
\end{equation}
The use of the real time instead of the parameterized one implies
that the random force autocorrelation function acquires a
different form:
\begin{equation}
                  \label{eq:XI_AVER_RH}
\left<\xi(t)\xi(t')\right>=k_{\rm B}\sqrt{T(t)T(t')}g(t)g(t')
\gamma_{\rm th}(t-t')=
k_{\rm B}\sqrt{T(t)T(t')}\gamma(t,t') \;.
\end{equation}
This FDR varies from Eq.~(\ref{eq:GFDR1}) insignificantly if
the temperature changes slowly during the solvent
response time, when $\gamma_{\rm th}(t-t')$ differs
from zero.
However,
the use of the effective time to describe the behavior of the Gaussian noise
in Eq.~(\ref{eq:FORCE1}) provides an important contribution
that allows the correct projected equations of motion to better
control the intensity of the stochastic force when the
bath is far from equilibrium.
This is the main difference between the Hamiltonian approach of the
current work and that of Ref.~\onlinecite{hern99c} where the random
force is governed only by the changing amplitude.

\subsection{Stochastic Dynamics}
\label{sec:stochastic}

The friction kernel of a free heavy particle, diffusing in solution
without external potentials ($V(q)=0$), has a
memory-less form,\cite{kneller04}
\begin{equation}
\gamma_{({\rm th})k}(\tau)=2\gamma_{0k}\delta(\tau)\;.
\end{equation}
Recalling the standard functional relation,
$
\delta(f(x))=\delta(x-x_0)/|f'(x_0)|\;,
$
where for simplicity we have
assumed that $x_0$ is the only root of the argument, {\it i.e.} $f(x_0)=0$,
then
Eqs.~(\ref{eq:FRICTION}) and (\ref{eq:XI_AVER}) lead to
\begin{eqnarray}
                  \label{eq:FRICTION_delta}
\gamma(t,t')&=&
\sum_k g_k^2(t)
\cdot 2\gamma_{0k}\delta(t-t') \equiv 2\eta(t)\delta(t-t') \;, \\
                  \label{eq:XI_AVER_delta}
\left<\xi(t)\xi(t')\right>&=&
k_{\rm B}\sum_k T_k(t)
g_k^2(t)\cdot 2\gamma_{0k}\delta(t-t') \equiv k_{\rm B}T_{\rm eff}(t)\gamma(t,t') \;,
\end{eqnarray}
where the effective temperature, $T_{\rm eff}(t)$, of the
Brownian particle is set to
\begin{equation}
    \label{eq:Teff_LE}
T_{\rm eff}(t)=\frac{\sum_k T_k(t)g_k^2(t)\gamma_{0k}}{\sum_k g_k^2(t)\gamma_{0k}}\;,
\end{equation}
so as to satisfy the FDR.

The effective temperature
is consistent with that of
Refs.~\onlinecite{kurchan00} and \onlinecite{kurchan05}
for the special case when the noise is memoryless
and
all the reservoirs are stable in time
---{\it i.e.}, when the temperatures $T_k$ are constant and $g_k=1$.
Herein, the resulting iGLE turns into the memoryless
irreversible Langevin equation (iLE):
\begin{subequations}\label{eq:DIFFUS_EQ}
\begin{eqnarray}
\ddot{q}(t)
      &=&-\eta(t)\dot{q}(t)+\xi(t)\;,\\
\left<\xi(t)\xi(t')\right>
      &=&
         2k_{\rm B}T_{\rm eff}(t)\eta(t)\delta(t-t') \;.
\end{eqnarray}
\end{subequations}
The form of this stochastic equation of motion is similar
to that of a particle diffusing in a medium with spatial
temperature gradients:~\cite{mahato96}
\begin{subequations}
\begin{eqnarray}
\ddot q
      &=&-\eta(q)\dot q+\xi(q,t)\;,\\
\langle \xi(q,t)\xi(q,t') \rangle
      &=&2k_{\rm B}T(q)\eta(q)\delta(t-t')\;.
\end{eqnarray}
\end{subequations}
In both cases, the particle quickly equilibrates as it traverses
from one local region to another, in conformity with the memoryless
limit.
Such a traversal is reminiscent of space-dependent friction dynamics
wherein the stochastic particle traverses between distinct
dissipating environments. The difference here is that the dissipation
can change even if the particles stays in the same region because of
changes in the dissipating environment.
Thus the earlier Darboux-like topological construction of the dissipating
environments is now extended to be time-dependent.


\subsubsection{Uniform temperature Limit}
\label{subsec:UniformT}

In the limit that all the reservoirs obey a homogeneous
temperature profile, $T_k(t)=T(t)$ for all $k$,
so that $T_{\rm eff}(t)=T(t)$ (see Eq.~(\ref{eq:Teff_LE})),
the friction coefficient in Eq.~(\ref{eq:DIFFUS_EQ}) becomes
\begin{equation}
\eta(t)=G^2(t)\gamma_0\;,
\end{equation}
where $G(t)$ is the effective coupling coefficient defined according to
Ref.~\onlinecite{hern06e},
\begin{equation}
G^2(t)=\sum_k \left. g^2_k(t)\gamma_{0k} \right/ \gamma_0
   \quad\mbox{and}\quad
\gamma_0=\sum_k \gamma_{0k}\,.
\end{equation}
We thus recover the iGLE of Ref.~\onlinecite{hern06e}
in which the stochastic particle is effectively coupled to a single
homogeneous (but time-dependent) reservoir.

\subsubsection{Two-Reservoir Limit}
\label{subsec:TwoTRes}

In Refs.~\onlinecite{MR95} and \onlinecite{ray98},
the LE for the Brownian particle is derived in
the case that the latter is dissipated by two reservoirs:
a global thermal bath at temperature $T$, and
a local nonequilibrium (time-dependent) bath.
The LE reads, as usual,
\begin{equation}
    \label{eq:NONEQ_LE}
\ddot{q}(t)=-(\eta_{\rm eq}+\eta_{\rm neq})\dot{q}(t)+\xi(t)\;,
\end{equation}
but with the stochastic force subdivided into equilibrium and nonequilibrium parts,
$\xi(t) = \xi_{\rm eq}(t) + \xi_{\rm neq}(t)$.
The FDR for the equilibrium part is
$\langle \xi_{\rm eq}(t)\xi_{\rm eq}(t') \rangle=2\eta_{\rm eq}k_{\rm B}T\delta(t-t')$.
The energy density of the nonequilibrium
bath modes changes with time and was found in
Refs.~\onlinecite{MR95} and \onlinecite{ray98} to be
\begin{subequations}\label{eq:NONEQ_U}
\begin{eqnarray}
u(\omega,t)&=&\frac{1}{4\eta_{\rm neq}}\int_{-\infty}^{\infty}d\tau
\langle\xi_{\rm neq}(t)\xi_{\rm neq}(t+\tau)\rangle e^{i\omega\tau}\\
&=& \frac{k_{\rm B}T}{2}+e^{-\nu t/2}\left[ u(\omega,0)-\frac{k_{\rm B}T}{2} \right]
\end{eqnarray}\end{subequations}
This takes into account the average dissipation
---$\nu$ in the notation of the earlier articles---
of the nonequilibrium reservoir modes due to their coupling
to the thermal reservoir.

Within the framework of the present approach,
all the modes
of the nonequilibrium reservoir have one and the same initial temperature,
$ u(\omega,0)=k_{\rm B}T_0/2 $.
The time-dependent temperature of this reservoir
according to Eq.~(\ref{eq:NONEQ_U}) is
\begin{equation}
    \label{eq:NONEQ_T}
T_{\rm neq}(t)= T+e^{-\nu t/2}( T_0-T )\;,
\end{equation}
and it is the same for all the modes.
Taking the inverse Fourier transform of Eq.~(\ref{eq:NONEQ_U}), one obtains
\begin{equation}
    \label{eq:NONEQ_XI}
\langle\xi(t)\xi(t+\tau)\rangle =
2k_{\rm B}T_{\rm eff}(t)\eta \delta(\tau)\;,
\end{equation}
where
\begin{equation}
    \label{eq:NONEQ_Teff}
T_{\rm eff}(t)=\frac{\eta_{\rm eq}T+
\eta_{\rm neq}T_{\rm neq}(t)}
{\eta_{\rm eq}+\eta_{\rm neq}}\,.
\end{equation}
This latter result is in agreement with the effective
temperature in Eq.~(\ref{eq:Teff_LE}) if $g_k(t)\equiv 1$.
The latter requirement is precisely what is needed to ensure that all
of the nonequilibrium baths reduce to the single local nonequilibrium
bath under consideration in this subsection.

Hence the current construction of the generalized effective temperature
of a particle connected to an arbitrary number of baths reduces to
the appropriate limit when it is connected to two nearly-separable
baths.

\subsubsection{Three-Reservoir Gedenken Experiments}\label{sec:simulations}
\label{subsec:ThreeTRes}

To further illustrate the accuracy of Eq.~(\ref{eq:Teff_LE}) in
providing the effective temperature of a subsystem interacting with
nonequilibrium reservoirs, we construct a simple
model for the diffusion of a heavy spherical particle,
with mass $M$, immersed in a three-component solvent gas of
light spherical particles.
The particles in each of the components
are assumed to have the same mass $m$.
Each component particle interacts only with other particles of the same component
through hard-sphere collisions so as to quickly achieve a
quasi-equilibrium temperature $T_k$ for the $k^{\rm th}$ component.
All the particles interact with the heavy particle
through hard-sphere collisions.
A critical assumption is that each component does not interact
with the other component, and thus each can maintain a distinct
temperature for very long times.
This severe assumption
is the origin of our use of the term ``gedenken'' to describe this model.

In principle, a numerical simulation of this system can be accomplished
using a many-particle Molecular Dynamics simulation in which each
solvent component does not in any way interact with each other.
However, if the desired observables are restricted to correlation
functions at times longer than the mean time to collision, then
these assumptions are sufficiently severe that the dynamics of
the system can instead be performed using a statistical approach
ala the Enskog theory for gases.
Briefly, the algorithm involves the free propagation of the
hard sphere particle in three-dimensional space
for a duration that extends to a randomized time that is
consistent with the statistics for collisions with each of the particles.
The collision takes place with a particle of type $k$ whose
3-dimensional momentum is consistent with a Boltzmann distribution
at temperature $T_k$.
After the collision, the heavy mass has a new momentum and proceeds
to the next collision.



More precisely, the time $t_k$ to the next collision between the heavy particle
and a particle from the $k$-th reservoir
is calculated as a Poissonian stochastic variable with the average value
\begin{equation}
        \label{eq:Collision_t}
\bar t_k = \frac{1}{ \pi (R+r_k)^2 \rho_k \overline u_k(\vec V)} \;,
\end{equation}
where $R$ is the radius of the solute particle,
$r_k$ is the radius of particle in the $k$-th reservoir,
and
$\rho_k$ is the number density of the $k$-th reservoir.
The mean relative velocity
$\overline u_k(\vec V) (\equiv \langle |\vec V - \vec v_k| \rangle)$
between the Brownian particle velocity $\vec V$
and the particle velocity $\vec v_k$ of the $k$-th reservoir
is given in Eq.~(\ref{eq:u_k}) below.
In the numerical simulation to be discussed, only three reservoirs
are included ---that is, $k\in\{1,2,3\}$--- but the algorithm would
still be efficient for an arbitrary number of reservoirs
as long as there are finite in number.
Given the stochastic times, $\bar t_1$, $\bar t_2$, and $\bar t_3$, the heavy particle
collides with the reservoir particle whose corresponding collision time
is the shortest.
The time $t^*$ and label $k^*$
of the closest collision corresponds to this minimal $\bar t_k$.

To calculate $\overline{u}_k$, we again invoke the assumption
that each $k^{\rm th}$ reservoir is distributed according to
a Boltzmann distribution that is unperturbed by its small
interaction with the heavy particle.\cite{yamaguchi01}
Thus, the Maxwellian
distribution of particles of the $k$-th reservoir,
\begin{equation}
        \label{eq:MAXWELL}
f_k(\vec v_k)=\left( \frac{m}{2\pi k_{\rm B}T_k} \right)^{3/2}
\exp\left( -\frac{mv_k^2}{2 k_{\rm B}T_k} \right) \;,
\end{equation}
leads to the average relative velocity,\cite{lowe98}
\begin{equation}
        \label{eq:u_k}
\overline u_k(\vec V) = \sqrt{\frac{2k_{\rm B}T_k}{\pi m}}
\exp \left( -\frac{mV^2}{2k_{\rm B}T_k}\right) +
\frac{k_{\rm B}T_k/m+V^2}{V}
{\rm erf}\left( V\sqrt{\frac{m}{2k_{\rm B}T_k}} \right)\;,
\end{equation}
where $V$ is the velocity of the heavy solute particle and the error function is
defined as ${\rm erf}(x)=2/\sqrt\pi\cdot \int_0^x \exp(-y^2)dy$.

Before the collision, the colliding bath particle is assigned a vector
$\vec v_{k^*}$ whose components are calculated from the corresponding Maxwellian
distribution~(\ref{eq:MAXWELL}).
After the collision, the change of the velocity of the heavy solute particle is
defined as
$$
d\vec V = \frac{m}{m+M}
\left( \vec v_{k^*} - \vec V + |\vec v_{k^*} - \vec V| \cdot \vec n \right)\;,
$$
where $\vec n$ is the unit scattering vector.
This vector is distributed isotropically when hard spheres
collide.\cite{LL1}

Three different gedenken simulations have been performed with the parameters
prescribed as follows.
The volume and the mass of the solute particle are taken
to be 64 times larger than that of
the reservoir particles which suffices to ensure the assumed time scale separation.
The specific values for the masses and radii are taken to be
$M=5.181\cdot 10^{-18}$~g, $m=8.095\cdot 10^{-20}$~g,
$R=4$~nm, and $r_k=1$~nm.
This is consistent with the known density of bulk gold
---{\it ca.~$19.32 \rm g/cm^3$}--- 
and corresponds to large and small spherical clusters with about 16,000 and 200
gold atoms, respectively. 
Such clusters have been readily constructed in the 
literature,\cite{dominey96,ogumi07}
though of course the separation of the light clusters into three
distinct reservoirs would be somewhat harder to maintain for 
extended times.

The number densities of the reservoirs are taken to be
$\rho_k=1.919\cdot 10^{-6} \; {\rm nm}^{-3}$ corresponding to a
dilute gas at pressures on the order of $10^{-4}\; \rm Atmosphere$.
The temperatures of the first and third reservoirs are $T_1=100$~K and
$T_3=500$~K, respective.
The temperature $T_2$ of the second reservoir is varied in each of the
simulations as noted in Table \ref{tab:temperatures}.
As the number densities of the three reservoirs are equal, their average temperature,
$T_{\rm avg}=(T_1+T_2+T_3)/3$, might
naively be expected to be the effective temperature of the heavy particle.
In order to not bias the results away from this naive estimate,
the heavy particles are initialized with velocities drawn from a
Maxwellian distribution at $T_{\rm avg}$.
We found that good convergence and adequate sampling for
the correlation functions of interest to this work
could been obtained for each case
by sampling 10,000 trajectories for up to 4ms.

The results of the simulations are presented in Table \ref{tab:temperatures},
where $\tau_V$ is the relaxation time of the velocity autocorrelation function
(VACF) of the heavy solute,
$\tau_E$ is the relaxation time of the solute's kinetic energy,
 and $T^{\rm sim}_{\rm eff}(\equiv\frac{M}{3k_{\rm B}}\langle V^2 \rangle)$
is the observed kinetic energy of the
Brownian particle expressed as an effective temperature.
\begin{table}
\begin{ruledtabular}
\begin{tabular}{ccccccccc}
& {\ \ \ }
& \multicolumn{3}{c}{relaxation time, $\mu$s}
& {\ \ \ }
& \multicolumn{3}{c}{temperature, K}
\\
\cline{3-5}\cline{7-9}
\; $T_2$, K \;
  && $\tau^{\rm sim}_V$
  & \;\; $\tau^{\rm theo}_V$
  & $2\tau_E$
&
  & \; $T^{\rm sim}_{\rm eff}$ \;
  & $T^{\rm theo}_{\rm eff}$
  & \; $T_{\rm avg}$ \;
\\
\hline
100 && \; 0.3513 & 0.3483 & 0.3621 \;
    &
    &    307.1 & 309.3 \;
    &    233.3\\
300 && \; 0.2984 & 0.2978 & 0.2992 \;
    &
    & 346.4 &348.6 \;
    & 300.0\\
500 && \; 0.2704 & 0.2703 & 0.2701 \;
    &
    & 422.3 &425.4 \;
    & 366.7\\
\end{tabular}
\end{ruledtabular}
\caption{The results of a numerical gedenken experiment
are summarized in this table.
The chosen particle is coupled to three distinct baths
with temperatures, $T_1$, $T_2$, and $T_3$, respectively.
In all three experiments,
$T_1 (=100{\rm K})$ and
$T_3 (=500{\rm K})$ are held constant while $T_2$ is varied
as noted in the table.
Various relaxation times and temperatures for all three experiments are
provided.
The ``sim" superscript denotes a result computed directly
from the simulation, and the ``theo" superscript denotes
a result calculated using the theory described in the text.
}
\label{tab:temperatures}
\end{table}
%
The theory discussed thus far predicts that $T_{\rm avg}$ will be equal
to $T_{\rm eff}$ only if all the coupling coefficients are equal.
Indeed, the corresponding temperatures listed in Table \ref{tab:temperatures}
are unequal indicating that a naive equilibrium treatment does
not suffice.

The effective temperature for these nonequilibrium systems can be calculated
with the help of the central expression from Eq.~(\ref{eq:central})
reduced to the form of Eq.~(\ref{eq:Teff_LE}) as appropriate for the
current model.
The friction coefficients for each reservoir are an input to this
equation, but they can readily be calculated in this case.
They are proportional to the average of
the reciprocal relaxation times of Eq.~(\ref{eq:Collision_t})
weighted by the velocity distribution $f(\vec V)$ of the heavy particle:
\begin{subequations}\label{eq:barbarUk}
\begin{equation}
        \label{eq:new_fric}
g_k^2\gamma_{0k}=\frac{2}{3} \pi (R+r_k)^2 \rho_k
        \overline{\overline{u}_k} \;,
\end{equation}
where
\begin{equation}
        \label{eq:detailbarUk}
\overline{\overline{u}_k}
  \equiv \langle\langle |\vec V - \vec v_k| \rangle\rangle =
         \int \overline{u}_k(\vec V) f(\vec V) d^3\vec V\;.
\end{equation}
\end{subequations}
If indeed the heavy mass motion can be described as having an effective
temperature $T_{\rm eff}$, then its velocity is simply the
the Maxwellian distribution at
the temperature $T_{\rm eff}$ that is to be determined.
Integration of Eq.~(\ref{eq:detailbarUk}) leads to
\begin{equation}
        \label{eq:bar_u_k}
\overline{\overline{u}_k}
= \int \overline u_k(\vec V) f(\vec V) d^3\vec V =
  \sqrt{\frac{8}{\pi}
  \left( \frac{k_{\rm B}T_{\rm eff}}{M} +\frac{k_{\rm B}T_k}{m} \right)}\;.
\end{equation}
Substitution of Eqs.~(\ref{eq:barbarUk}) and (\ref{eq:bar_u_k})
in (\ref{eq:Teff_LE}) gives the self-consistent equation,
\begin{equation}
        \label{eq:Teff_self_consist}
T_{\rm eff}=
\frac{\sum_k \sqrt{T_{\rm eff}/M+T_k/m}\cdot T_k}
     {\sum_k \sqrt{T_{\rm eff}/M+T_k/m}} \,.
\end{equation}
The effective temperatures $T_{\rm eff}^{\rm theo}$ calculated
from this theory either by solving this self-consistent equation or
by entering the inverse of the observed relaxation times into the
basic Equation (\ref{eq:Teff_LE}) agree within the digits of accuracy shown
in Table \ref{tab:temperatures} and are thus included as a single
value for each gedenken experiment.
The good agreement between the effective temperatures
predicted by the nonequilibrium theory and the simulated effective
temperatures demonstrates that the coupling coefficients
play a role in determining the apparent temperature of a probe
particle in the nonequilibrium situation that it is
in contact with distinct reservoirs of varying temperatures.


The VACF relaxation time $\tau_V$ can also be determined using the arguments
employed thus far.
It takes the value,
\begin{equation}
        \label{eq:tau_V_theo}
\tau^{-1}_V=
\sum_k g_k^2\gamma_{0k} =
\frac{2}{3}\sum_k \pi (R+r_k)^2 \rho_k \overline{\overline{u}_k}\;,
\end{equation}
which can be readily computed using the results from
Eqs.~(\ref{eq:barbarUk}) and (\ref{eq:bar_u_k}).
The values are listed in the table as $\tau_V^{\rm theo}$.
Again, the agreement between the nonequilibrium theory
and the simulation is remarkable.
As a check on the quality of the simulations, the relaxation time
$\tau_E$ in the kinetic energy of the heavy particle was also obtained
by simulation.
The expected result form the theory of Brownian motion is that
$2\tau_E$ will be equal to the velocity relaxation time as is indeed
observed within nominal error bars.


\subsubsection{Local Quasi-Equilibrium Limit}
\label{subsec:LocalQuasiEq}

In Ref.~\onlinecite{rubi04}, slowly relaxing systems
---effectively at local quasi-equilibrium---
are described on the basis of Onsager's fluctuation
theory.
Therein, the
non-Markovian Fokker-Planck equation for the conditional probability
density, $P(\vec{\alpha},t;\vec{\alpha}_0,t_0)$, was shown to be
\begin{equation}
    \label{eq:FPqe}
\frac{\partial P}{\partial t} = \sum_{i,j}
\frac{\partial }{\partial \alpha_i}B_{ij}(\vec{\alpha},t,t_0)
\left[ X_j(\vec{\alpha},t)P +
\frac{k_{\rm B}T(t)}{m} \frac{\partial P}{\partial \alpha_j} \right]
\;,
\end{equation}
where $m$ is the particle's mass;
$\vec{\alpha}=\{\alpha_i$\} is the set of fluctuating variables and
$\vec{\alpha}_0$ are their values at time $t_0$.
The thermodynamic forces, $X_i$, are defined through the quasi-equilibrium
probability density $P_{\rm qe}$ as follows:
\begin{equation}
X_i(\vec{\alpha},t)= -\frac{\partial }{\partial \alpha_i}
\left[
\frac{k_{\rm B}T(t)}{m} \ln P_{\rm qe}(\vec{\alpha},t)\right]
\;.
\end{equation}
The coefficients, $B_{ij}$,
define the connection between the stream velocities,
$v_{\alpha_i}$, and the conjugate thermodynamic forces,
\begin{equation}
v_{\alpha_i} = -\sum_j B_{ij}\left[ X_j + \frac{k_{\rm B}T(t)}{m}
\frac{\partial}{\partial \alpha_j} \ln P \right] \;,
\end{equation}
in accordance with the rules of nonequilibrium thermodynamics.
The temperature $T(t)$
of the system is assumed to be time-dependent, but at a slow enough
rate that the system achieves local quasi-equilibrium.
This temperature is a function only of time and is expressed through that of the bath,
$T_{\rm B}(t)$, as
\begin{equation}
T(t)=A(t)T_{\rm B}(t) \;.
\end{equation}
The factor $A(t)$ is defined {\it via} the thermodynamic functions taking into
account the energy exchange
between the system and the bath and is equal to unity when the process is reversible.

Note that the factors $A(t)$ and $B_{ij}(\vec{\alpha},t,t_0)$ are not specified
within the approach of Santamaria-Honeck et al,\cite{rubi04}
and other methods must be used to obtain them.
Thus, the Brownian motion in a granular gas has been investigated
using a kinetic theory to calculate these factors.\cite{santos99}
In this case, the behavior of a particle is described through its velocity, so
that the stochastic variable has only one component, $\alpha_1 \equiv \dot q$.
The factor $A$ has been found to be constant and identified through the
restitution coefficient $\lambda$ as $A=(1+\lambda)/2$,
and the single matrix
element $B_{11}=A\gamma(t,t_0)$, where $\gamma(t,t_0)$ is the
friction coefficient which depends only on time.

The quasiequilibrium distribution is
\begin{equation}
P_{\rm qe}\sim \exp\left[ -\frac{m\dot q^2}{2k_{\rm B}T(t)} \right] \;,
\end{equation}
and the Fokker-Planck equation~(\ref{eq:FPqe}) takes the form
\begin{equation}
    \label{eq:FPqe1}
\frac{\partial P}{\partial t} =
\frac{\partial }{\partial \dot q} \left[ B_{11}(t,t_0)\dot q P \right] +
\frac{\partial^2 }{\partial \dot q^2} \left[ B_{11}(t,t_0)
\frac{k_{\rm B}T(t)}{m} P \right] \;.
\end{equation}
It corresponds to the iLE of the general form~(\ref{eq:DIFFUS_EQ})
\begin{equation}
\ddot q = -B_{11}(t,t_0)\dot q + \xi(t) \;,
\end{equation}
where the stochastic force correlation function is
\begin{equation}
\left<\xi(t)\xi(t')\right> = 2\frac{B_{11}(t,t_0)}{m}
k_{\rm B}T(t)\delta(t-t')\;.
\end{equation}
Thus we, once again, find agreement between the current framework
and earlier work.

\section{Conclusion}
           \label{sec:conclusion}

In this article, we have presented a generalized construction
for the effective temperature of a tagged particle connected to
an arbitrary number of time-dependent inhomogeneous reservoirs.
It is in agreement with several limiting cases described earlier
by various authors.\cite{hern06e,mahato96,kurchan05,MR95,ray98,rubi04}
In Ref.~\onlinecite{hern06e}, it was shown by comparison between theory
and simulations that the Brownian diffusion of a tagged particle
(or probe) within swelling hard spheres at constant temperature
can be surmised by an iLE.
The latter is the memoryless limit of the
iGLE, when the memory kernel $\gamma(t,t')$ is proportional to the
$\delta$-function, $\delta(t-t')$.
The formalism presented here
is suitable for describing situations
when the environment solvating a tagged particle is itself
out of equilibrium.
The particle, thus, ``experiences"
different environments with differing properties
(say, temperature and viscosity) as it moves in time.

The numerical simulation of a particle diffusing through a
gas that is somehow composed of three distinct temperature
particle baths discussed in Sec.~\ref{sec:simulations} is instructive.
Clearly, in the limit that the three baths can couple to each
other, they will equilibrate to the same average temperature.
The chosen particle will likewise attain this equilibrium temperature
which is also equal to the effective temperature defined by
Eq.~(\ref{eq:central}) in this limit.
The numerical gedenken experiment provides the result for the opposite limit
in which the three baths are completely uncoupled to each other.
The perhaps surprising result to a naive
observer, who incorrectly assumes that this is an
equilibrium process, is that the chosen particle does not
exhibit dynamics at the average temperature of the three baths.
Instead, the degree of coupling between the baths and the
chosen particle modulates the heat transfer that goes
between the distinct baths through the particle.
The effective temperature exhibited by the chosen particle,
as given by Eq.~(\ref{eq:central}), is precisely the mathematical form
for this delicate balance.
The existence of energy flows through the chosen particle
would in the long time limit lead to the re-equilibration of
all the baths.
However, when the baths are large enough, such energy transfer is sufficiently
small that it doesn't change the temperatures of the baths.
Thus the presence of a subsystem at an effective temperature
unequal to the average temperature of its nonequilibrium surroundings
will be seen whenever
the re-equilibration time of the surroundings is much longer than the
time scales of interest in the subsystem.

Such behavior should be seen in solution chemistry in which
the solvent molecules undergo chemical reactions,
leading to a substantial (and possibly heterogeneous)
change in the solvation of the solutes.
The differences between these neighborhoods depend strongly on the
differences between the reactant and product solvent molecules.
Moreover, the energetics of the reactions may also lead to localized
energy losses or gains that would manifest themselves as temperature
fluctuations.
Although each of the solvent reservoirs would not be as simply
decomposable as in the gedenken simulations of Sec.\ref{sec:simulations},
they would nevertheless interact with the solute heterogeneously
leading to an effective temperature as given by Eq.~(\ref{eq:central}).


Another illustration lies in the nonequilibrium dynamics of colloidal microgel
particles.
The latter can change their volume in response
to the temperature or pH alteration of the solution.\cite{lyon01}
In accordance with recent unpublished experiments by Lyon,\cite{lyon_PC}
these colloids, being initially in a glassy state,
can form a liquid after decreasing their size or
crystallize when swelling.
The generalization of the iGLE to incorporate temporal temperature changes
thus extends our earlier theory\cite{hern06e}
to account for the diffusion of particles
in temperature-dependent colloidal suspensions.




\section{Acknowledgments}%

This work has been partially supported by the
National Science Foundation through Grant
Nos.~NSF 02-123320 and 04-43564.
RH is the Goizueta Foundation Junior Professor.

\appendix
\section{}     \label{app:A}

In this appendix, we sketch the derivation of
the solution of Eq.~(\ref{eq:MODE})
quoted in the text as Eq.~(\ref{eq:Xi}).
For simplicity, all the indices associated with the various baths
are omitted, and consequently Eq.~(\ref{eq:MODE})
takes the form:
\begin{equation}
\ddot x + \nu(t) \dot x + \Omega^2(t) x = c
\frac{g'(t)h(t)}{\mu^2(t)}q
\;.
\end{equation}
After the replacement,
\begin{equation}
                  \label{eq:REPLACE1}
x=uw\;,
\end{equation}
the previous equation becomes
\begin{equation}
                  \label{eq:UV_E}
\ddot w u + 2 \dot u \dot w + \nu(t) \dot w u + w [\ddot
u+\nu(t)\dot u+\Omega^2(t) u]= \frac{c g'(t)h(t)}{\mu^2(t)} q \;,
\end{equation}
which greatly simplifies ---{\it cf.}~Eq.~(\ref{eq:V_E})---
if only $u$ satisfies the auxiliary differential equation,
\begin{equation}
                  \label{eq:U_E}
\ddot u+\nu(t)\dot u+\Omega^2(t) u = 0\;,
\end{equation}
for the initial conditions,
\begin{equation}
                  \label{eq:U_INIT}
u(0) = 1 \;, \ \dot u(0) =0 \;.
\end{equation}

The auxiliary function $u$ in Eq.~\ref{eq:U_E})-(\ref{eq:U_INIT})
can be obtained using the steepest descent approximation.
The usual substitution,
\begin{equation}
                  \label{eq:WKB0}
u=\exp(S)\;,
\end{equation}
leads to
\begin{equation}
                  \label{eq:WKB}
{\dot S}^2+\Omega^2(t)=-\nu(t)\dot S - \ddot S \;.
\end{equation}
Since the right hand side (RHS) is a small quantity ($\nu$ and $\ddot S$ vanish
if $\mu$ and $h$ are constant),
the zeroth-order approximation is determined entirely by $\Omega$:
\begin{equation}
\dot S_0=\pm i\Omega(t)\;.
\end{equation}
Substituting $\dot S_0$ into the RHS of Eq.~(\ref{eq:WKB}), we obtain
a differential equation for the first-order correction, $S_1$,
\begin{equation}
                  \label{eq:WKB1}
{\dot S_1}^2+\Omega^2(t)=-\nu(t)\dot S_0 - \ddot S_0 \;.
\end{equation}
Using the definitions~(\ref{eq:MODE_COEFF}) of $\nu$ and $\Omega$,
the solution for $\dot S_1$ can be written, after some algebra, as
\begin{equation}
\dot S_1(t)=\pm i\Omega(t)\sqrt{1\pm \frac{i}{\Omega(t)}\frac{d}{dt}
\ln\left( \mu(t)h(t) \right)};.
\end{equation}
In the case of an adiabatic change of parameters $\mu(t)$ and $h(t)$,
the inequality
\begin{equation}
     \label{eq:Acond}
\left| \frac{d}{dt}\ln\left( \mu(t)h(t) \right) \right| =
\left| \frac{\dot\mu}{\mu}+\frac{\dot h}{h} \right| \ll \Omega(t)
\end{equation}
is satisfied.
Therefore,
\begin{equation}
\dot S_1(t) \approx \pm i\Omega(t) - \frac{1}{2}\frac{d}{dt}
\ln\left(\mu(t) h(t)\right)\;,
\end{equation}
and
\begin{equation}
S_1(t) \approx \pm i\int_0^t\Omega(t')dt' - \ln\sqrt{\mu(t) h(t)}\;.
\label{eq:S1TWO}
\end{equation}

The two complex solutions for $u$ ---each found by substitution of
Eq.~(\ref{eq:WKB0}) by Eq.~\ref{eq:S1TWO}---
can now be combined to ensure that the initial conditions in
Eq.~(\ref{eq:U_INIT}) are satisfied, leading to
\begin{equation}
                  \label{eq:U}
u(t)\approx\frac{\cos \omega\tau(t)}{\sqrt{\mu(t)h(t)}} \;,
\end{equation}
where
\begin{equation}
\tau(t) \equiv \frac{1}{\omega}\int_0^t{\Omega(t')dt'}=\int_0^t{\frac{h(t')}{\mu(t')}dt'} \;,
\end{equation}
and the approximation is satisfied according to the
inequality in Eq.~(\ref{eq:Acond}).
This now leads to the desired intermediate differential
equation in the unknown auxiliary function $w$,
\begin{equation}
                  \label{eq:V_E}
\ddot w + 2 \frac{\dot u(t)}{u(t)} \dot w + \nu(t) \dot w =
\frac{cg'(t)h(t)}{u(t)\mu^2(t)} q(t)
\end{equation}
with the initial conditions,
\begin{equation}
                  \label{eq:V_INIT}
w(0) = x(0) \;, \ \dot w(0) = \dot x(0) \;,
\end{equation}
which follow from Eqs.~(\ref{eq:REPLACE1}) and (\ref{eq:U_INIT}).


Eq.~(\ref{eq:V_E}) can be manipulated further using the
substitution,
\begin{equation}
\dot w\equiv ab \;,
\end{equation}
leading to a new differential equation,
\begin{equation}
                  \label{eq:AB_E}
\dot a b + a \left[\dot b+\left(\nu(t)+2\frac{\dot u}{u}\right)
b\right]= \frac{c g'(t)h(t)}{u(t)\mu^2(t)} q(t) \;.
\end{equation}
This expression simplifies if only $b$ is chosen to satisfy
the auxiliary differential equation,
\begin{equation}
                  \label{eq:B_E}
\dot b+\left(\nu(t)+2\frac{\dot u(t)}{u(t)}\right) b = 0 \;,
\end{equation}
which gives rise to the solution,
\begin{equation}
                  \label{eq:B}
b(t) = \frac{b(0)}{(u(t)\mu(t))^2}\;.
\end{equation}
After substitution, Eq.~(\ref{eq:AB_E}) reduces to
\begin{equation}
                  \label{eq:A_E}
\dot a = \frac{cg'(t)h(t)}{b(0)}u(t)q(t) \;.
\end{equation}

The solution of Eq.~(\ref{eq:A_E}) is
\begin{equation}
                  \label{eq:A}
a(t) = a(0) + \frac{c}{b(0)}\int_0^t{g'(s)h(s)u(s)q(s)ds} \;.
\end{equation}
The result for $w$ can now be obtained by
multiplying Eqs.~(\ref{eq:B}) and (\ref{eq:A}),
leading to
\begin{equation}
                  \label{eq:DOT_V}
\dot w = a(t)b(t) =\frac{a(0)b(0)}{(u(t)\mu(t))^2}
+ \frac{c}{(u(t)\mu(t))^2}\int_0^t{g'(s)h(s)u(s)q(s)ds}
\;.
\end{equation}
From the initial conditions in Eq.~(\ref{eq:V_INIT}), it follows that
$\dot x(0)=\dot w(0)=a(0)b(0)$. Hence,
\begin{equation}
                  \label{eq:V}
w(t) = x(0) + \dot x(0)\int_0^t\frac{dt'}{(u(t')\mu(t'))^2} +
\int_0^t \frac{c\,dt'}{(u(t')\mu(t'))^2}\int_0^{t'}{g'(s)h(s)u(s)q(s)ds}
\;.
\end{equation}
The desired solution of Eq.~(\ref{eq:MODE}) quoted in the
primary text as Eq~(\ref{eq:Xi}) now
follows form
Eqs.~(\ref{eq:REPLACE1}), (\ref{eq:U}) and (\ref{eq:V}).

\section{}     \label{app:B}

In this appendix, we rederive the iGLE projection shown
earlier in Ref.~\onlinecite{hern99e}
for the extended case of multiple disconnected
baths that is of interest to this work.

Substitution of Eq.~(\ref{eq:Xi}) into Eq.~(\ref{eq:MOTION4}) gives
\begin{subequations} \label{eq:QQ}
\begin{eqnarray}
                  \label{eq:Q}
\ddot q &=& \sum_{\bf i} c_{\bf i}g'_k(t)h_k(t)u_{\bf i}(t)x_{\bf i}(0) +
\sum_{\bf i} c_{\bf i}g'_k(t)h_k(t)u_{\bf i}(t)\dot x_{\bf i}(0) \int_0^t
 \frac{dt'}{(u_{\bf i}(t')\mu_k(t'))^2} \\
                  \label{eq:Q1}
&+& \sum_{\bf i} c^2_{\bf i}g'_k(t)h_k(t)u_{\bf i}(t) \int_0^t
\frac{dt'}{(u_{\bf i}(t')\mu_k(t'))^2}\int_0^{t'}
g'_k(s)h_k(s)u_{\bf i}(s)q(s)ds\\
&-& \frac{\partial V(q)}{\partial q} - \sum_{\bf i}
\left(\frac{c_{\bf i}g'_k(t)}{\omega_{\bf i}} \right)^2 q -
\frac{\partial}{\partial q}\delta V_2(q(\cdot),t)     
\end{eqnarray}
\end{subequations}
where we use the notation ${\bf i}\equiv (i,k)$ as in the text,
and the summation over $\bf i$ includes all the bath modes in
all the reservoirs, {\it i.e.}, $\sum_{\bf i}\equiv\sum_k\sum_i$).
The second line~[Eq.~(\ref{eq:Q1})]
can be rewritten as
\begin{equation}
                  \label{eq:X1}
X= \sum_{\bf i} c_{\bf i}^2g'_k(t)h_k(t)u_{\bf i}(t) \int_0^t
g'_k(s)h_k(s)u_{\bf i}(s)q(s)ds
\int_{s}^{t}\frac{dt'}{(u_{\bf i}(t')\mu_k(t'))^2} \;
\end{equation}
after a change in the order of integration, {\it i.e.}, by noting that
$\int_0^tdt'\int_0^{t'}ds=\int_0^tds\int_{s}^tdt'$.
The inner integral
---appearing also in the first line~(\ref{eq:Q})---
can be found readily with the help of the substitution
$d\tau_k'=\left(h_k(t')/\mu_k(t')\right)dt'$
[{\it cf.}~Eq.~(\ref{eq:TAU})]:
\begin{equation}
                  \label{eq:X1_INT}
\int_{s}^{t}  \frac{dt'}{(u_{\bf i}(t')\mu_k(t'))^2} =
\int_{s}^{t}  \frac{h_k(t')/\mu_k(t')}{\cos^2(\omega_{\bf i}\tau_k')}dt' =
\int_{s}^{t}  \frac{d\tau_k'}{\cos^2(\omega_{\bf i}\tau_k')} =
\frac{1}{\omega_{\bf i}}\left(\tan \omega_{\bf i}\tau_k(t)
 - \tan \omega_{\bf i}\tau_k(s)\right)
\end{equation}
Insertion of this result in Eq.~(\ref{eq:X1})
and substitution of $u_{\bf i}(t)$ according to Eq.~(\ref{eq:Ui}) gives
\begin{eqnarray}
                  \label{eq:X2}
X&=& \sum_{\bf i} c_{\bf i}^2 g'_k(t)h_k(t)
 \frac{\cos \omega_{\bf i}\tau_k(t)}{\sqrt{\mu_k(t)h_k(t)}}\nonumber\\
 & \cdot &
 \int_0^t g'_k(s)h_k(s)\frac{\cos \omega_{\bf i}\tau_k(s)}{\sqrt{\mu_k(s)h_k(s)}}
 q(s)\frac{\tan \omega_{\bf i}\tau_k(t) - \tan \omega_{\bf i}\tau_k(s)}{\omega_{\bf i}}ds\\
&=& \sum_{\bf i} \frac{c_{\bf i}^2g_k(t)}{\omega_{\bf i}}\frac{h_k(t)}{\mu_k(t)}
 \int_0^t g_k(s)\frac{h_k(s)}{\mu_k(s)} q(s)
 \sin \omega_{\bf i}(\tau_k(t)-\tau_k(s)) ds \nonumber \\
&=& \sum_{\bf i} \frac{c_{\bf i}^2h_k(t)}{\omega_{\bf i}^2\mu_k(t)}g_k(t)
 \int_0^t g_k(s) q(s) d\cos \omega_{\bf i}(\tau_k(t)-\tau_k(s))
\;, \nonumber
\end{eqnarray}
where
$g'_k(t)$ has also been expanded according to Eq.~(\ref{eq:new_g}).
Integration by parts, leads to
\begin{eqnarray}
                  \label{eq:X3}
X &=&-\sum_{\bf i}\frac{c_{\bf i}^2h_k(t)}{\omega_{\bf i}^2\mu_k(t)}g_k(t)
 \int_0^t \cos \omega_{\bf i}(\tau_k(t)-\tau_k(s)) g_k(s) \dot q(s) ds
 +\sum_{\bf i}\left( \frac{c_{\bf i}g'_k(t)}{\omega_{\bf i}} \right)^2 q \\
&-&
\sum_{\bf i}\frac{c_{\bf i}^2h_k(t)}{\omega_{\bf i}^2 \mu_k(t)} g_k(t) q(0)
 \cos\omega_{\bf i}\tau_k(t) -
\sum_{\bf i}\frac{c_{\bf i}^2h_k(t)}{\omega_{\bf i}^2 \mu_k(t)} g_k(t)
\int_0^t \cos \omega_{\bf i}(\tau_k(t)-\tau_k(s)) \dot g_k(s) q(s) ds
\;. \nonumber
\end{eqnarray}

Substitution of the results
of Eqs.~(\ref{eq:X1_INT}) and (\ref{eq:X3}) into the differential
equation (\ref{eq:QQ}) for $q$ leads to
\begin{eqnarray}
             \nonumber
\ddot q &=& -\frac{\partial V(q)}{\partial q} -\int_0^t \gamma(t,t')\dot q(t')dt' + \xi(t) \\
        &-& \sum_{\bf i}\frac{c_{\bf i}^2h_k(t)}{\omega_{\bf i}^2 \mu_k(t)} g_k(t)
            \int_0^t \cos \omega_{\bf i}(\tau_k(t)-\tau_k(s)) \dot g_k(s) q(s) ds
                  \label{eq:Q_APP_B}
            -\frac{\partial}{\partial q(t)}\delta V_2(q(\cdot),t) \\
                  \label{eq:FRICTION_APP_B}
\gamma(t,t')&=&
\sum_{\bf i} \frac{c_{\bf i}^2}{\omega_{\bf i}^2}\frac{h_k(t)}{\mu_k(t)}
g_k(t)g_k(t')\cos\omega_{\bf i}(\tau_k(t)-\tau_k(t')) \;, \\
                  \label{eq:FORCE_APP_B}
\xi(t)&=&\sum_{\bf i} \frac{c_{\bf i}}{\omega_{\bf i}}\frac{h_k(t)}{\mu_k(t)}g_k(t)
\left(
p_{\bf i}(0)\sin\omega_{\bf i}\tau_k+
\left( \omega_{\bf i}x_{\bf i}(0)-\frac{c_{\bf i}}{\omega_{\bf i}}q(0) \right)
 \cos\omega_{\bf i}\tau_k
\right) \;.
\end{eqnarray}
Taking the derivative $\partial(\delta V_2)/\partial q$
with the help of the Euler-Lagrange variational principle and
ignoring the higher order contributions from
the nonstationary memory correction, as suggested by
Ref.~\onlinecite{hern06e},
one sees that
this derivative coincides with the formal differentiation of Eq.~(\ref{eq:V2_0})
in $q(t)$,
\begin{equation}
\frac{\partial}{\partial q(t)}\delta V_2 = - \int_0^t dt' a(t,t') q(t')  \;,
\end{equation}
with $a(t,t')$ from Eq.~(\ref{eq:A_NONLOC}), and
two last terms in Eq.~(\ref{eq:Q_APP_B}) cancel each other.



\bibliography{j,hern,flucbar,mfpt,liquid,noneq,surf,gas,glass,bio,tst,md}
\bibliographystyle{apsrev}
\end{document}